\documentclass[twocolumn]{aastex62}

\newcommand{\QDeb}{\textsc{qdeblend${}^{\mathrm{3D}}$}}
\received{27 September 2018}
\revised{18 April 2019}
\accepted{23 May 2019}
\submitjournal{ApJ}
\begin{document}

\title{Jet-driven galaxy-scale gas outflows in the hyper-luminous quasar 3C~273}

\correspondingauthor{Bernd Husemann}
\email{husemann@mpia.de}

\author[0000-0002-0786-7307]{B. Husemann}
\affiliation{Max-Planck-Institut f\"ur Astronomie, K\"onigstuhl 17, 69117 Heidelberg, Germany}

\author[0000-0003-2064-0518]{V. N. Bennert}
\affiliation{Physics Department, California Polytechnic State University, San Luis Obispo CA 93407, USA}

\author[0000-0003-3804-2137]{K. Jahnke}
\affiliation{Max-Planck-Institut f\"ur Astronomie, K\"onigstuhl 17, 69117 Heidelberg, Germany}

\author[0000-0003-4932-9379]{T. A. Davis}
\affiliation{School of Physics \& Astronomy, Cardiff University, Queens Buildings, The Parade, Cardiff, CF24 3AA, UK}

\author[0000-0002-8055-5465]{J.-H. Woo}
\affiliation{Department of Physics and Astronomy, Seoul National University, Seoul 151-742, Republic of Korea \\
}

\author[0000-0003-1585-9486]{J. Scharw\"achter}
\affiliation{Gemini Observatory, Northern Operations Center, 670 N. A'ohoku Place, Hilo, HI 96720, USA}

\author[0000-0002-6660-6131]{A. Schulze}
\affiliation{National Astronomical Observatory of Japan, Mitaka, Tokyo 181-8588, Japan}

\author[0000-0003-2754-9258]{M. Gaspari}
\altaffiliation{\textit{Lyman Spitzer Jr.} Fellow}
\affiliation{Department of Astrophysical Sciences, Princeton University, 4 Ivy Lane, Princeton, NJ
08544-1001, USA}

\author[0000-0003-0101-1804]{M. Zwaan}
\affiliation{European Southern Observatory, Karl-Schwarzschild-Str. 2, 85748 Garching b. M\"unchen, Germany}


\begin{abstract}
We present an unprecedented view on the morphology and kinematics of the extended narrow-line region (ENLR) and molecular gas around the prototypical hyper-luminous quasar 3C~273 ($L_\mathrm{bol}\sim10^{47}\mathrm{erg\,s}^{-1}$ at $z=0.158$) based on VLT-MUSE optical 3D spectroscopy and ALMA observations. We find that: 1) The ENLR size of $12.1\pm0.2$kpc implies a smooth continuation of the size-luminosity relation out to large radii or a much larger break radius as previously proposed. 2) The kinematically disturbed ionized gas with line splits reaching 1000\,$\mathrm{km\,s}^{-1}$ out to $6.1\pm1.5$kpc is aligned along the jet axis. 3) The extreme line broadening on kpc scales is caused by spatial and spectral blending of many distinct gas clouds separated on sub-arcsecond scales with different line-of-sight velocities. The ENLR velocity field combined with the known jet orientation rule out a simple scenario of a radiatively-driven radial expansion of the outflow. Instead we propose that a pressurized expanding hot gas cocoon created by the radio jet is impacting on an inclined gas disk leading to transverse and/or backflow motion with respect to our line-of-sight. The molecular gas morphology may either be explained by a density wave at the front of the outflow expanding along the jet direction as predicted by positive feedback scenario or the cold gas may be trapped in a stellar over-density caused by a recent merger event. Using 3C~273 as a template for observations of high-redshift hyper-luminous AGN reveals that large-scale ENLRs and kpc scale outflows may often be missed due to the brightness of the nuclei and the limited sensitivity of current near-IR instrumentation. 
\end{abstract}

\keywords{ISM: jets and outflows --- quasars: individual: 3C 273 --- quasars: emission lines --- techniques: imaging spectroscopy}

\section{Introduction} \label{sec:intro}
The quasi-stellar radio source (quasar) 3C~273 was found to have a redshift of $z\sim0.158$ \citep{Schmidt:1963}, which revealed that quasars are of extra-galactic origin and cannot be stars. 3C~273 is one of the most luminous quasars at low redshift with an absolute brightness of $M_V=-26.8$\,mag \citep[e.g.][]{Hamilton:2008} and a radio flux of $f_\mathrm{1.4GHz}\sim45$\,Jy \citep{Kellermann:1969} corresponding to $L_{\mathrm{1.4GHz}}=2.9\times10^{27}\,\mathrm{W\,Hz}^{-1}$. 3C~273 is therefore a proto-typical active galactic nucleus (AGN) since only accretion by a super-massive black hole (SMBH) can produce the observed energy \citep{Salpeter:1964}. 

The enormous energy released by luminous AGN like 3C~273 is suspected to significantly affect the evolution of their host galaxies either through radiative AGN feedback \citep[e.g][]{Silk:1998} or mechanical feedback via their jets \citep[e.g.][]{Pedlar:1990}. While the impact of  jets on the hot halo gas around massive galaxies has already been confirmed \citep[see][for a review]{Fabian:2012}, the capability of AGN radiation to accelerate and expel a large gas reservoir from AGN host galaxies is currently under intense investigation. 

Large reservoirs of ionized gas on tens to hundreds kpcs scales have been detected around many luminous radio-loud and radio-quiet AGN at low redshifts \citep{Stockton:1983,Danziger:1984,vanBreugel:1985,Stockton:1987,Prieto:1993,Shopbell:1999,Tadhunter:2000, Stockton:2002, Villar-Martin:2005, Villar-Martin:2010, Husemann:2010, Husemann:2011, Greene:2012, Liu:2013, Liu:2014, Hainline:2014, Villar-Martin:2017, Villar-Martin:2018} as well as at high redshifts \citep[e.g.][]{Heckman:1991b,McCarthy:1993,Christensen:2006,North:2012,Cantalupo:2014,Hennawi:2015,Borisova:2016, Cai:2017, ArrigoniBattaia:2018,Husemann:2018}. These are often referred to as extended emission-line regions (EELRs) or extended narrow-line regions (ENLRs) in the literature without a clear distinction between the two. However, the largest ionized nebulae have predominantly been associated with radio-loud AGN hosting large-scale jets. The size and morphology of EELRs are therefore often interpreted as being redistributed material due to shock fronts created by the expanding radio jet \citep[e.g.][]{Tadhunter:2000,Tremblay:2018} or a signature of tidal debris from galaxy interactions \citep[e.g.][]{Villar-Martin:2010,Villar-Martin:2018}. In this paper, we use the term EELR for ionized gas on 10 to several 100\,kpc scales independent of the ionization mechanism and the term ENLR only for clearly AGN ionized gas out to several tens of kpc around luminous AGN.

Independent of the size of the ionized nebulae, the prevalence of kpc-scale outflows around AGN remains a key question to understand AGN feedback and its impact on host galaxies. Spectroscopic observations have therefore been crucial to map fast AGN-driven outflow from low-luminosity  \citep[e.g.][]{Storchi-Bergmann:1992,Crenshaw:2000,Riffel:2011,Fischer:2013,Venturi:2018} to high-luminosity AGN \citep[e.g.][]{Greene:2011,Liu:2013b,Harrison:2014,Carniani:2015,McElroy:2015,Kakkad:2016,Karouzos:2016} and AGN with powerful radio jets leading to jet-cloud interactions \citep[e.g.][]{Tadhunter:1989,Emonts:2005,Holt:2008,Morganti:2013,Mahony:2016,Villar-Martin:2017,Nesvadba:2017,Santoro:2018,Jarvis:2019}.  Yet, the relative role of jets and AGN radiation as the main driver for fast outflows and their diverse properties among the overall AGN population still needs to be investigated. 

3C~273 represents an ideal laboratory to study the relative impact on the radiation and the jet with the surrounding medium. While 3C~273's radio jet has been studied in detail over the past 50 years, observations of its host galaxy remain difficult due to the overwhelmingly bright nucleus. The host galaxy was first detected from the ground by \citet{Wyckoff:1980} and followed up with \textit{Hubble} at high-angular resolution \citep{Bahcall:1995b,Martel:2003}. \citet{Wyckoff:1980} and \citet{Boroson:1985} already noted evidence of extended emission lines a few arcsec away from the nucleus, but \citet{Stockton:1987} did not detect an EELR in [\ion{O}{3}] $\lambda5007$ using narrow-band imaging. \citet{Hippelein:1996} presented the first map of the ENLR from Fabry-Perot imaging.

In this article, we present a combined study of deep optical integral-field unit (IFU) observations of 3C~273 with the Multi-Unit Spectroscopic Explorer \citep[MUSE,][]{Bacon:2010} and CO(1-0) mapping with archival ALMA observations.  The MUSE data reveal highly complex ionized gas motions in the host galaxy of 3C~273 and the ALMA observations highlight substantial amounts of molecular gas. We discuss our observations in the light of an expanding hot gas cocoon inflated by the radio jet which creates a fast outflow and shock front in the ISM. Furthermore, we explore the detectability of similar ENLRs around hyper-luminous ($L_\mathrm{bol}>10^{47}\,\mathrm{erg\,s}^{-1}$) AGN at high-redshift as studied for example by the WISSH project \citep{Bischetti:2017}. Throughout the paper we assume a concordance flat $\Lambda$CDM cosmology with $H_0=70\,\mathrm{km\,s}^{-1}\,\mathrm{Mpc}^{-1}$, $\Omega_m=0.3$ and $\Omega_\Lambda=0.7$ corresponding to 2.73kpc/\arcsec\ at the redshift of 3C~273.

\begin{figure*}
\includegraphics[width=\textwidth]{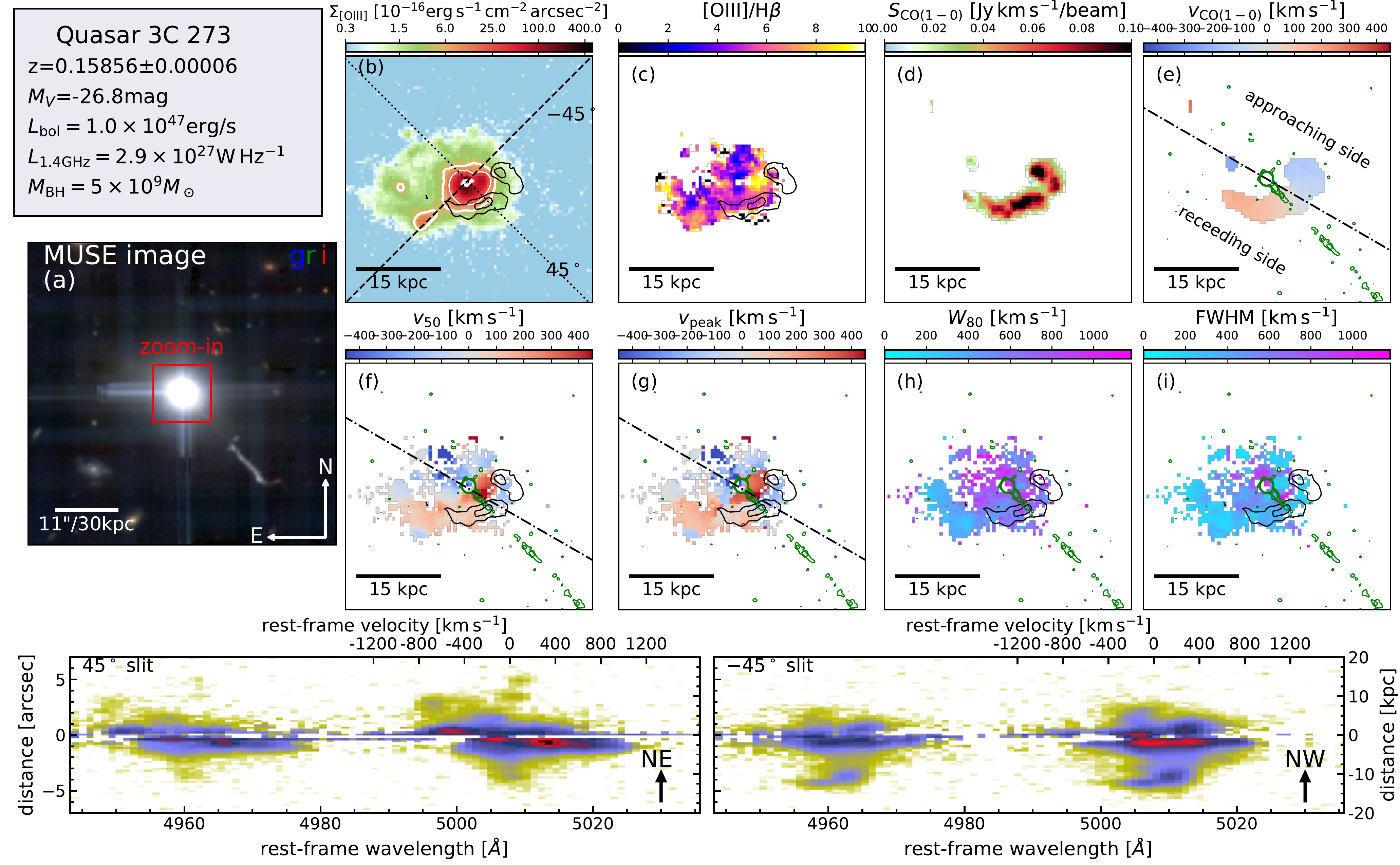}
 \caption{Overview of the ionized and molecular gas properties in 3C~273 from MUSE and ALMA observation. \textit{(a)} Re-constructed $gri$ RGB-color image from the MUSE observation covering a $1\arcmin\times1\arcmin$ FoV. (b)  [\ion{O}{3}] $\lambda5007$ narrow-band image after PSF and continuum subtraction in the zoom-in region. The white contours refer to surface brightness limits of $1\times$ and $3\times10^{-15}\,\mathrm{erg}\,\mathrm{s}^{-1}\mathrm{cm}^{-2}\mathrm{arcsec}^{-2}(1+z)^{-4}$. (c) [\ion{O}{3}]/H$\beta$ emission line ratio map with CO(1-0) distribution shown as contours. (d) CO(1-0) flux map as recovered from the ALMA observations. (e) CO(1-0) cold gas velocity field from the 1st moment maps. The dot-dashed line divides the kinematic map into the receding and approaching side at zero velocity going through the QSO center. Green contours show the X-band (10\,GHz) radio map at $0\farcs25$ resolution from \citet{Perley:2017}. (f), (g), (h), (i) show the median velocity ($v_{50}$), the peak velocity ($v_\mathrm{peak}$), the $W_{80}$ line width and the FWHM of the [\ion{O}{3}] line profile, respectively. The CO(1-0) distribution is shown as black contours in all four panels as a reference.   \textit{Bottom panels:} Re-constructed long-slit spectra  of the [\ion{O}{3}] $\lambda4960,5007$ lines after PSF and continuum subtraction along the two slit axes as indicated in panel (b).}
 \label{fig:3C273_overview}
\end{figure*}

\section{Observations \& Analysis} \label{sec:observations}
\subsection{MUSE observations}
MUSE data of 3C~273 were obtained under ESO program 097.B-0080(A) on 31 March 2016 under gray moon, clear sky conditions and a seeing of 0\farcs7. The data cover a $1'\times1'$ field-of-view (FoV) at $0\farcs2$ sampling and a wavelength range of 4750--9300\AA\ at a spectral resolution of $R\sim2500$. Given the brightness of 3C~273 ($m_V=12.6$mag), we split the observations into  $18\times250$s exposures with a total integration time of 4500s. Small sub-arcsec dithering and standard $90^\circ$ rotations between exposures were applied to mitigate residuals of the flat-fielding. The data were fully reduced and calibrated using the standard ESO pipeline \citep{Weilbacher:2014}.

The reconstructed $gri$ MUSE color image is shown in Fig.~\ref{fig:3C273_overview} (panel a). The bright AGN point source and the optical jet of 3C~273 are both well visible. An important processing step is therefore the deblending of the unresolved AGN emission from the spatially-resolved emission of the host galaxy. We use the software \QDeb\ \citep{Husemann:2013a,Husemann:2014} to estimate the point-spread function from the 2D intensity distribution of the H$\delta$, H$\beta$, \ion{He}{1}, and H$\alpha$  broad line wings originating from the unresolved AGN broad-line region. Afterwards we interpolate the wavelength-dependent PSF with a 2nd order polynomial on a spaxel-by-spaxel basis. The AGN spectrum from the central spaxels is then convolved with the PSF and subtracted from the original data. Since the H$\alpha$ line is saturated at the peak of the PSF, we cannot use the central spaxel for the wavelength range around H$\alpha$. Hence, we rather use a region north-east of the AGN which shows a deficit in extended emission to create a ``pure'' AGN spectrum around H$\alpha$ which we convolve again with the corresponding PSF and subtract it from the original data. Given the brightness of the AGN we still obtain a very high S/N spectrum of the AGN a few arcsec away from the center which does not leave strong residuals after subtraction. Nevertheless, the spaxels with $<$0\farcs8 around the AGN position remain corrupted in H$\alpha$ due to the saturation and need to be ignored completely.

\begin{figure*}
\includegraphics[width=\hsize]{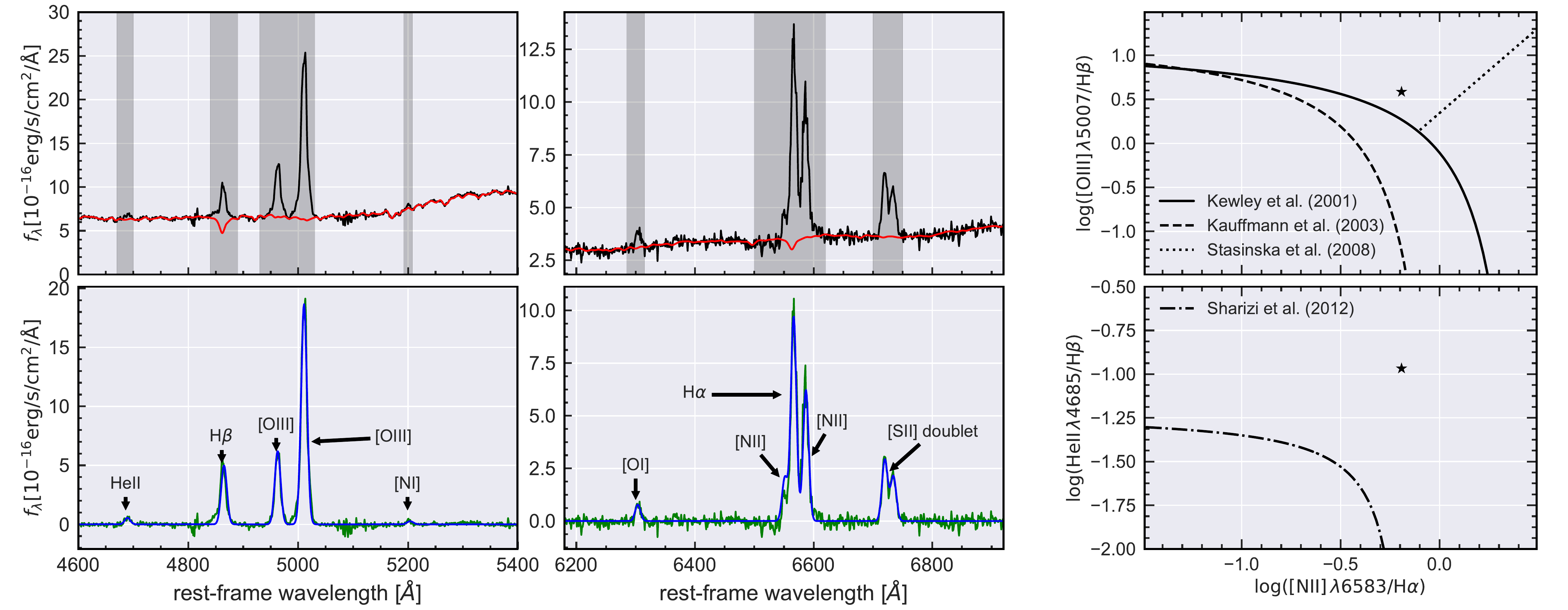} 
\caption{Aperture spectra for the H$\beta$ (left panels) and H$\alpha$ spectral range (middle panels) which are extracted from a radius of 4\arcsec\ after subtracting the point-like nucleus. The central 0\farcs8 was excluded due to high continuum residuals after subtracting the bright nucleus (in case of H$\beta$) and saturation of H$\alpha$ in the central spaxels. \textit{Top panels:} Integrated spectrum (black) and best-fit stellar continuum modelling (red). Regions with emission lines excluded from the continuum model are shown as dark shaded areas. \textit{Bottom panels:} Residual spectrum after subtracting the best-fit stellar continuum model (green) and the best-fit model of the emission lines (blue). All emission lines in this wavelength region are annotated. \textit{Right panels:} Emission-line diagnostic diagrams based on the \ion{He}{2}, H$\beta$, [\ion{O}{3}], H$\alpha$, and [\ion{N}{2}] lines measured in the integrated spectra. Corresponding demarcation lines from \citet{Kewley:2001}, \citet{Kauffmann:2003}, \citet{Stasinska:2008} and \citet{Shirazi:2012}. In both diagrams the measured line ratios fall in the regime of AGN ionization.}\label{fig:continuum}
\end{figure*}

As shown in Fig.~\ref{fig:3C273_overview}, we clearly recover spatially extended [\ion{O}{3}] emission with complex kinematics even at an AGN-host galaxy contrast ratio of about 200 in the continuum. In addition, we also recover the stellar continuum of the host galaxy at high S/N  within an annulus of 0\farcs8--6\arcsec circular radii (Fig.~\ref{fig:continuum} top panel). We modelled the stellar continuum by fitting stellar templates from the INDO U.S. library \citep{Valdes:2004} using the \textsc{PyParadise} software \citep{Walcher:2015,Weaver:2018}. From the stellar continuum fit we infer an accurate systemic redshift of $z=0.15850\pm0.00005$ and a stellar velocity dispersion within this larger aperture of $\sigma_*=210\pm10\,\mathrm{km\,s}^{-1}$. The integrated spectrum reveals line ratios of [\ion{O}{3}\,$\lambda5007$]/H$\beta=3.9\pm0.1$, [\ion{N}{2}\,$\lambda6583$]/H$\alpha=0.64\pm0.04$, and \ion{He}{2}\,$\lambda4685$/H$\beta=0.11\pm0.01$, where \ion{He}{2}\,$\lambda4685$ is detected with 10$\sigma$ confidence. We show the corresponding emission-line diagnostic diagrams in Fig.~\ref{fig:continuum} together with classical demarcation lines, which clearly show that star formation is not the dominant ionization mechanism of the ionized gas. Due to the complex kinematics and heavy blending of the [\ion{N}{2}] and H$\alpha$ lines we cannot robustly map this line ratio across the nebula. However, the [\ion{O}{3}\,$\lambda5007$]/H$\beta$ line ratio map can be constructed (Fig.~\ref{fig:3C273_overview} panel c) and reveals that the line ratio is even higher in most parts confirming that AGN and/or shock ionization is the main ionization mechanisms across the nebula.

\subsection{ALMA observations}
ALMA uses 3C~273 as a calibrator so that a large number of short observations of this source are available. We retrieved all data available for 3C~273 covering the red-shifted frequency of $^{12}$CO(1-0) with a spatial resolution between 0\farcs5 and 1\farcs0. This included 8 different tracks from 5 different ALMA projects, all of which observed 3C~273 for $\approx$5 minutes in order to use it as a bandpass calibrator. In all but one case a 2\,GHz correlator window (with a raw spectral resolution of 15.625 MHz) was present over the redshifted frequency of CO(1-0). In one case a 1850\,MHz correlator window was present, with a raw spectral resolution of 3.906 MHz. The other spectral windows present in each observation were used to detect continuum emission.

The raw ALMA data for each track were calibrated using the standard ALMA pipeline in the \texttt{Common Astronomy Software Applications} {(\tt CASA)} package. Additional flagging was carried out where necessary to improve the data quality. Three iterations of self calibration were used to improve the phase and amplitude calibration of the data on the bright continuum in 3C~273. The data presented here were produced using natural weighting, yielding a synthesized beam of 0\farcs78\,$\times$\,0\farcs75 at a position angle of 37$^{\circ}$ (a physical resolution of $\approx$2.1\,kpc). We use data with a channel width of 30\,km\,s$^{-1}$, and pixels of 0\farcs2 (resulting in approximately 3.5 pixels across the synthesized beam).

Bright point source continuum emission with a flux of 12.9$\pm$0.02\,Jy was detected over the full line-free bandwidth. Based on the position of this point source we are able to register the relative AGN positions of the MUSE and ALMA data with a precision of at least one pixel size (0\farcs2). We subtracted the continuum emission from the data in the $uv$ plane using the {\tt CASA} task {\tt uvcontsub}. The continuum-subtracted dirty cubes were cleaned in regions of source emission (identified interactively) to a threshold equal to 1.5 times the RMS noise of the dirty channels.  The clean components were then added back and re-convolved using a Gaussian beam of full-width-at-half-maximum (FWHM) equal to that of the dirty beam.  This produced the final, reduced, and fully calibrated $^{12}$CO(1--0) data cube of 3C~273, with a RMS noise levels 0.18\,mJy beam$^{-1}$ in each 30\,km\,s$^{-1}$ channel.
 
CO(1-0) emission was detected from this source, with an integrated line width of $\approx$490\,km\,s$^{-1}$ (FWHM), and an integrated intensity of 1.82$\pm$0.02\,Jy\,km\,s$^{-1}$. This corresponds to an H$_2$ mass of (1.7$\pm$0.3)$\times10^9$\,M$_{\odot}$ assuming an $\alpha_{\rm CO}$ = 0.8\,M$_{\odot}$ (K\,km\,s$^{-1}$pc$^2$)$^{-1}$ as typically used for compact luminous systems, such as starburst galaxies, SMGs and QSOs \citep[e.g.][]{Downes:1998, Scoville:2003,Xia:2012}. If we were instead to use a Galactic $\alpha_{CO}$ then the derived H$_2$ mass in this system would increase by a factor of 5.5.
 
Zeroth moment (integrated intensity), first moment (mean velocity), and second moment (velocity dispersion) maps of the detected line emission were created using a masked moment technique. A copy of the clean data cube was first Gaussian-smoothed spatially (with a FWHM equal to that of the synthesized beam), and then Hanning-smoothed in velocity. A three-dimensional mask was then defined by selecting all pixels above a fixed flux threshold of 1.5\,$\sigma$, adjusted to recover as much flux as possible in the moment maps while minimizing the noise.  The moment maps were then created using the un-smoothed cubes within the masked regions only. In Fig.~\ref{fig:3C273_overview} we show the zeroth and first moment maps in panel (d) and (e), respectively.

The morphology of the molecular gas is surprisingly asymmetric and appears as an ``arc''-like feature about $\sim1\farcs8$ away from the nucleus towards the south-west almost exactly perpendicular to the jet axis. It seems consistent with the dust lane identified in coronagraphic images presented by \citet{Martel:2003}, but the significantly lower spatial resolution of the ALMA data does not allow a precise spatial comparison. Importantly, the molecular arc is significantly more extended along the south-west direction on both sides of the spiral continuum excess structure seen in the \textit{Hubble} images. In particular, there is no corresponding feature in \textit{Hubble} images at the location of the brightest molecular gas knot seen 2\arcsec\ west of the nucleus. Since the ionized gas morphology is completely different and does not show a similarly prominent structure it is unclear whether the molecular gas is bound to a stellar over-density formed in a recent galaxy interaction or is currently forming in-situ due to a density wave and enhanced ambient pressure initiated by the expanding outflow.

\begin{figure*}
 \includegraphics[width=\hsize]{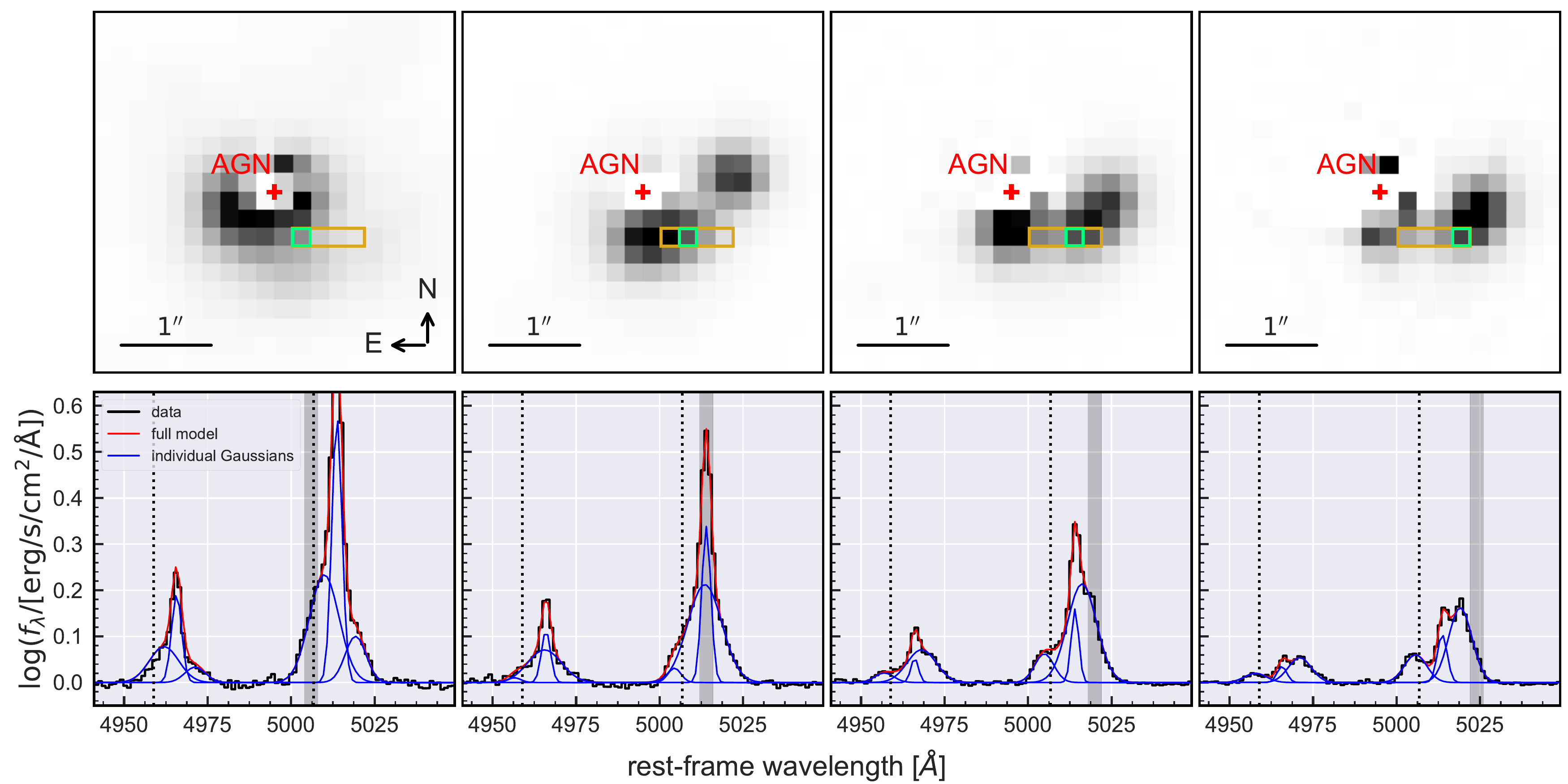}
 \caption{Zooming in on the complex nature of the [\ion{O}{3}] line shape. We show the spectra of four adjacent spaxels in the MUSE cube in the bottom panels. The position of the 4 adjacent spectra are marked by the green squares in the associated continuum-subtracted narrow-band images (top panels). Those images are 4\,\AA\ wide bands with the corresponding wavelength ranges centered on the four dominant line components as highlighted by the dark grey shaded areas. The [\ion{O}{3}] spectrum shape changes significantly from spaxel to spaxel with sizes of 0\farcs2$\times$0\farcs2, which is about a factor of 3 smaller than the FWHM of the seeing.}
 \label{fig:O3_line}
\end{figure*}

\subsection{Multi-phase gas kinematics}
With MUSE we recover extended [\ion{O}{3}] emission out to $\sim$20\,kpc from the quasar. The [\ion{O}{3}] emission is asymmetric towards the east side as shown in Fig.~\ref{fig:3C273_overview} panel (b). The [\ion{O}{3}] surface brightness distribution confirms the results of the Fabry-Perot imaging \citep{Hippelein:1996}, but MUSE achieves a much higher S/N and covers a larger velocity range. In the bottom panels of Fig.~\ref{fig:3C273_overview}, we present two pseudo-slit spectra (slit width of 0\farcs8) for two position angles, one along the radio jet axis (PA $-45^\circ$ N to E) and one perpendicular to it (PA $45^\circ$). In both cases we see highly complex ionized gas motions, with amplitudes of $\pm800$\,$\mathrm{km\,s}^{-1}$ as well as several broad and multiply-peaked lines over this large velocity range in the central 4\,kpc. 

Figure~\ref{fig:O3_line} highlights the complexity of the [\ion{O}{3}] line profiles which are difficult to model with a fixed number of Gaussians. It is highly desired to dissect the [\ion{O}{3}] line profile into distinct components, which has been used to study individual physical conditions by correlating line ratios and kinematics \citep[e.g.][]{McElroy:2015}. We have tried this process but find that fitting [\ion{O}{3}] line profiles separately for each spaxel violates the spatial correlation function imposed by the PSF as shown in Fig.~\ref{fig:O3_line}. The [\ion{O}{3}] line shape is significantly changing from spaxel to spaxel which have a size of 0\farcs2, about a factor of 3 smaller than the FWHM of the seeing. While 3 Gaussians provide a good fit to the overall line profile in all cases, the spatial variation of the Gaussian parameters are unphysical as the PSF demands that individual spectral components only vary in flux on such small scales. A fully consistent solution would only be achieved when the line shapes are modelled directly in 3D taking the PSF into account as a constraint on the spatial flux distribution. Such a method, which is able to model the entire ENLR structure at once with the large number of components, still needs to be developed. For practical reasons, we rather map non-parametric line shape parameters such as the median velocity ($v_{50}$), the peak velocity ($v_\mathrm{peak}$), the FWHM and the width containing 80\% of the line flux ($W_{80}$), which has been previously used to to characterize ionized gas outflows in the [\ion{O}{3}] line around luminous AGN \citep[e.g.][]{Harrison:2014,Liu:2013b,McElroy:2015,Harrison:2016b,Sun:2017}. Since the method requires that the cumulative line flux distribution is monotonically increasing, i.e. no negative flux values across the line shape, we still need to generate a noise free model of the lines for each spaxel. Given the complex line shapes, we use a large spectral library of [\ion{O}{3}] doublet lines to model the line profiles as a non-negative linear superposition which is not restricting the model to a fixed number of Gaussians. The library is constructed with a velocity range of $\pm1200\,\mathrm{km\,s}^{-1}$ at $10\,\mathrm{km\,s}^{-1}$ sampling and four line dispersions ($40\,\mathrm{km\,s}^{-1}$, $80\,\mathrm{km\,s}^{-1}$, $160\,\mathrm{km\,s}^{-1}$, $320\,\mathrm{km\,s}^{-1}$), which are convolved with the instrumental resolution as inferred by \citet{Bacon:2017} at the [\ion{O}{3}] wavelength. During the fitting we discard components with a peak flux density S/N$<$3 to avoid fitting noise features. We use a Monte Carlo approach to estimate errors for each parameter  by re-fitting the noise-modulated data 50 times. In Fig.~\ref{fig:3C273_overview} panel f, g, h, and i we show the resulting kinematics maps with a $v_{50}$ velocity error less than $50\,\mathrm{km\,s}^{-1}$ and S/N$>$5 for $W_{80}$ and FWHM.

The kinematics maps inferred from [\ion{O}{3}] as shown in Fig.~\ref{fig:3C273_overview} are complex which are therefore best interpreted in conjunction with the molecular gas distribution and kinematics. The molecular gas corresponds to kinematically quiescent gas with a small line dispersion ($\leq45\,\mathrm{km\,s}^{-1}$) and a smooth velocity gradient from blue to red-shifted motion of $\pm160\,\mathrm{km\,s}^{-1}$ around the systemic velocity of the stellar body. The ionized gas shows the same smooth velocity gradient and radial velocity amplitude in $v_{50}$ and $v_\mathrm{peak}$ of $\pm150\,\mathrm{km\,s}^{-1}$ at the location of the molecular gas with a small line width both in $W_{80}$ and FWHM. Furthermore, the ionized gas at even larger distance from the AGN exhibits the same direction of motion as the molecular gas which suggest that the large scale kinematics are dominated by gravitationally-driven rotational motion. These quiescent motions at radii $>2\arcsec$ define the global rotational pattern with red-shifted motion on the east and blue-shifted motion on the west side of the galaxy.

On the contrary, the ionized gas kinematics enclosed within the radius of the molecular gas arc is highly complex as demonstrated in Fig.~\ref{fig:O3_line}. Nevertheless, the light-weighted bulk motion of the gas shows a bipolar velocity gradient along the radio jet axis with kinematics predominately red-shifted on the approaching jet side and blue-shifted on the opposite side. While the FWHM remains narrow with $<$400\,$\mathrm{km\,s}^{-1}$ across most of the ENLR it appears larger at the transition to the rotation-dominated region. This is likely a superposition effect due to the PSF smearing at the transition region. Likewise, the large $W_{80}>800\,\mathrm{km\,s}^{-1}$ width is not caused by individual clouds with high velocity dispersion but rather the spatial blending of many narrow emission lines with a broad distribution in radial velocities. While this complexity limits our ability to understand the interaction with ambient medium in detail, a key observation is that the radio jet axis is clearly aligned with the bipolar velocity gradient in the ionized gas indicative of a strong coupling between the expanding radio jet and the ambient gas in 3C~273. Although it is strange that the approaching side of the jet appears to push the gas backwards, we will outline a potential geometrical scenario in Sect.~\ref{sec:wind_results} based on the directly inferred relative orientation of the jet and gas disc.

\begin{figure*}
 \includegraphics[width=\hsize]{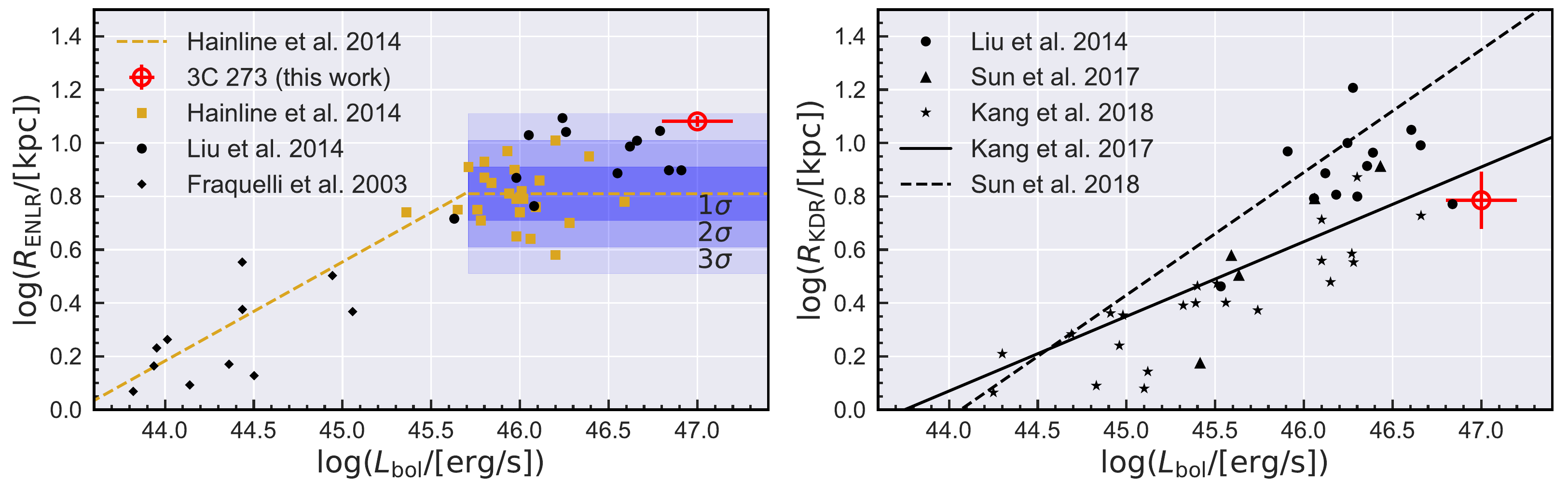}
 \caption{Scaling relation of ENLR size (left panel) and KDR size (right panel) against AGN luminosity. Literature data and proposed scaling relations are available mainly for obscured AGN and taken from various works for comparison purposes \citep{Fraquelli:2003,Hainline:2014,Liu:2014,Sun:2017,Kang:2018}. If necessary IR and [\ion{O}{3}] luminosities were converted to bolometric luminosities adopting $L_\mathrm{bol}/L_{8\mu\mathrm{m}}\sim10$ and $L_\mathrm{bol}/L_\mathrm{[OIII]}\sim3200$ of \citet{Runnoe:2012} and \citet{Pennell:2017}, respectively. The blue shaded areas in the left panel correspond to the 1$\sigma$, 2$\sigma$, and 3$\sigma$ deviations of the data from \citet{Hainline:2014} around the proposed ENLR break radius.} \label{fig:ENLR_scaling}
\end{figure*}

\section{Results \& Discussion} \label{sec:results}
\subsection{ENLR size and kinematically disturbed region}\label{sec:ENLR_results}
A lot of different definitions have been adopted for the ENLR size, such as an effective radius \citep[e.g.][]{Husemann:2014,Kang:2018}, the maximum detectable size \citep{Bennert:2002, Villar-Martin:2010,Greene:2012,Villar-Martin:2018,Storchi-Bergmann:2018} or the size till an intrinsic surface brightness limit corrected for cosmological dimming \citep[e.g.][]{Hainline:2013,Liu:2013,Hainline:2014,Sun:2017}. By design all those definitions are prone to different observational effects and their application is limited to specific science goals. The maximum detectable [\ion{O}{3}] emitting region size is strongly dependent on the depth of observations and whether the ionization mechanisms can be ambiguously attributed to the AGN ionization. It is therefore not well suited to study scaling relations between AGN luminosity and the ENLR size. Additionally, the luminosity-weighted radius depends on whether the dominating NLR is included or excluded from the radius estimations.

Several studies therefore adopted a characteristic ENLR size using a fixed intrinsic [\ion{O}{3}] line flux surface brightness limit of $10^{-15}\,\mathrm{erg}\,\mathrm{s}^{-1}\,\mathrm{cm}^{-2}\mathrm{arcsec}^{-2}(1+z)^{-4}$ to allow to construct a comparable ENLR size--luminosity relation at various redshifts. For 3C~273 we determine a characteristic ENLR size of $R_\mathrm{ENLR}=12.1\pm0.2$\,kpc, which is robust given the good spatial resolution and the application of the PSF subtraction technique. In Fig.~\ref{fig:ENLR_scaling} (left panel) we compare this ENLR to literature measurements. Previously, a break in the $L_\mathrm{AGN}$--$R_\mathrm{ENLR}$ scaling relation has been reported to occur somewhere at $\sim$6--10\,kpc \citep[e.g.][]{Liu:2013,Hainline:2013,Hainline:2014}. In order to mitigate the effect of beam smearing for their high-redshift targets, \citet{Hainline:2014} applied a surface brightness profile fitting convolved with an approximate PSF to recover the intrinsic size of the ENLR. Considering that we were able to subtract the dominating AGN point source directly and that we can exclude a top-hat function as a radial light ENLR profile we compare our inferred ENLR size for 3C~273 with the Sersic profile measurements from \citet{Hainline:2014}. We find that the ENLR size of 3C~273 is $2.7\sigma$ larger than the proposed break radius and sample distribution reported by \citet{Hainline:2014}. This suggests that either the ENLR size--luminosity relation has no break radius at all, that the break in the ENLR size occurs at significantly larger radii, or that the ENLR is generically asymmetric at the chosen surface brightness limit so that the sizes are systematically underestimated in randomly orientied long-slit observations. Given that 3C~273 is an unobscured AGN, the intrinsic ENLR size may be even larger compared to the obscured AGN used in previous studies because the unified model predicts that the ionization cone of unobscured AGN should be more aligned towards the line-of-sight than for obscured AGN. This would mean that statistically a large correction factor needs to be applied to turn the measured projected ENLR sizes into intrinsic sizes. We can rule out that the ionization cone is directly aligned with our line of sight to explain the relatively round morphology, because the kinematic substructure and the known direction of the radio jet implies a significant inclination of the cone. In any case, those inclination effects would only strengthen our result as the intrinsic size of the ENLR would be even higher for 3C~273. Interestingly, we note that adopting  a $3\times$ higher intrinsic [\ion{O}{3}] surface brightness limit as proposed by \citet{Sun:2018} would imply a size of $R_\mathrm{NLR}=4.15\pm0.1$\,kpc which is a factor 4 smaller than the corresponding $L_\mathrm{AGN}$--$R_\mathrm{ENLR}$ scaling relation reported by the authors. Hence, the choice of the surface brightness limit to measure the size has a great impact on the signatures of a ENLR break radius.  Additionally, unaccounted systematic effects in the relation of \citet{Sun:2018} may contribute to this discrepancy due to the use of broad-band imaging that still require spectroscopic confirmation.

A break in the size-luminosity relation would necessarily imply a transition from an ionization-bounded to matter-bounded region, e.g. running out of gas, as the radiation field smoothly decreased as $R^{-2}$. Hence, a transition may occur when the gas density drops faster than the radiation field at the edge of galaxies \citep{Netzer:2004,Hainline:2013}. The stellar body of 3C~273 can be traced out to a radius of 7\arcsec\ (19\,kpc) with an effective radius of $r_\mathrm{eff}=2.6\arcsec$\ \citep{Martel:2003}. Hence, the ENLR does not reach beyond the host galaxy and is well covered within $3\times r_\mathrm{eff}$ of the stellar distribution. A break in the ENLR size--luminosity relation may not occur at the chosen surface brightness limit simply because of the large size of 3C~273's host galaxy. Alternatively, \citet{Dempsey:2018} proposed that the transition from ionization- to matter-bounded occurs because optically-thick clouds lead to a steeper decline of the radiation field than $R^{-2}$ causing a break in the relation at small ENLR sizes. For 3C~273 we find the dense molecular gas to be located inside the characteristic ENLR radius so that dense gas shielding and filtering of AGN radiation may be responsible for the asymmetry of the ENLR and possibly reducing the ENLR brightness along the south-west direction but not towards the east side. Another scenario relevant for 3C~273 is the cooling of gas from the hot halo gas. According to the chaotic cold accretion (CCA) scenario, ionized gas filaments and clouds are condensing out of the hot phase within $\sim$10kpc \citep[][their Fig. 5]{Gaspari:2018}. Due to CCA, the transition from ionization to matter-bounded ENLR clouds would be related to the cooling radius where a large portion of gas becomes neutral again to be ionized by the AGN radiation. While a fraction of the condensed gas will be funneled towards the SMBH, a residual part will settle onto a clumpy rotating structure within the central 10\,kpc scale, which is in agreement with our observations.

Overall, none of the processes that have been proposed to reduce the ENLR break radius seems relevant for 3C~273. This is in agreement with our result that 3C~273 is inconsistent with the reported break radius in the ENLR size--luminosity relation. However, very large EELRs have been reported around luminous AGN at low surface brightness levels \citep[e.g.][]{Villar-Martin:2010,Husemann:2011,Kreimeyer:2013,Villar-Martin:2018} order of magnitude below the intrinsic surface brightness assumed for the characteristic ENLR radius here. Usually those EELR structures are in a matter-bounded regime far beyond the transition radius \citep[e.g.][]{Fu:2007,Kreimeyer:2013,Storchi-Bergmann:2018} and are therefore usually associated with gas density enhancements in the intergalactic medium due to tidal tails, ongoing galaxy interactions or ram pressure from an expanding radio jet. For 3C~273 we do not detect any [\ion{O}{3}] emission beyond 23\,kpc at our surface brightness detection limit of $\Sigma_{\mathrm{[OIII]}}>4\times10^{-17}\,\mathrm{erg\,s}^{-1}\mathrm{cm}^{-2}\mathrm{arcsec}^{-2}$ at $3\sigma$ level per 0\farcs2 pixel.  This suggests that 3C~273 has not been subject to a gas rich merger very recently and the expanding radio jet has not expelled gas beyond the 20\,kpc scale.

\begin{figure*}
 \includegraphics[width=\hsize]{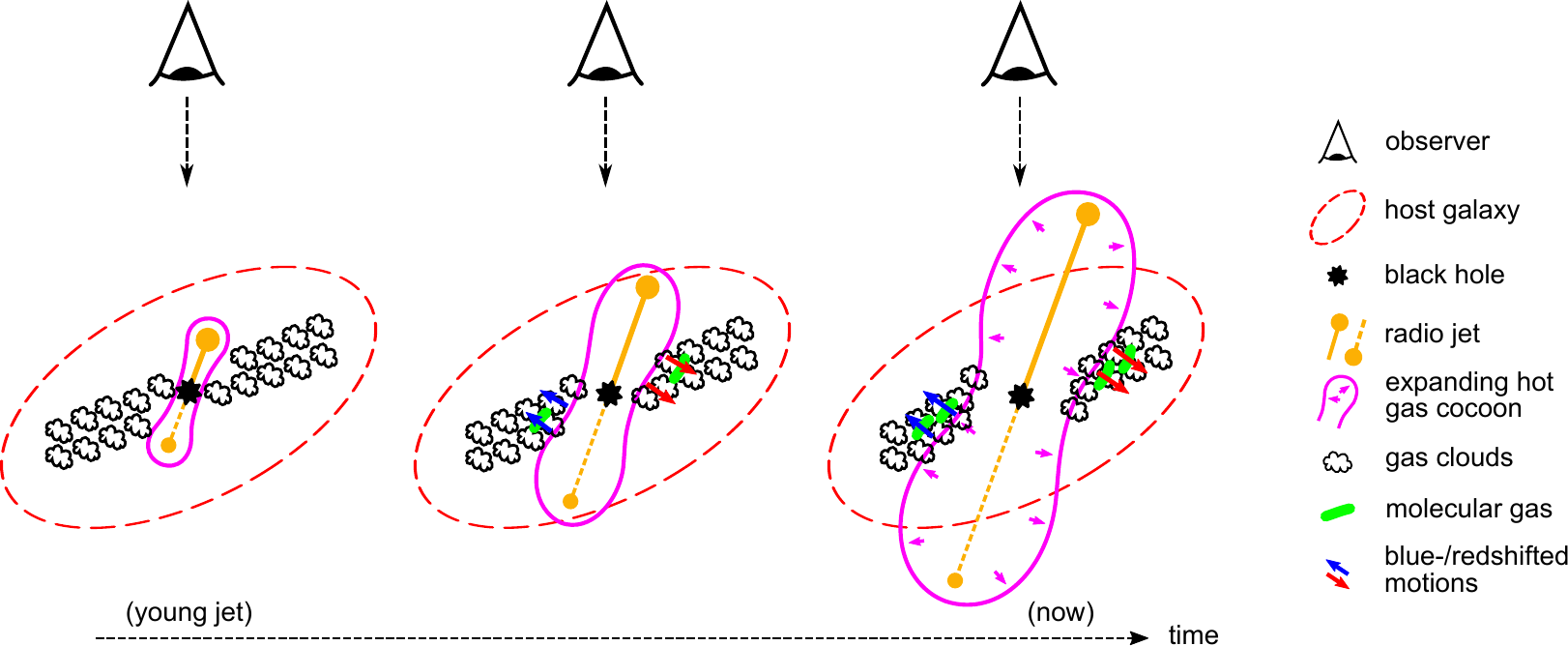}
 \caption{A cartoon describing our proposed scenario to explain the unusual gas kinematics of 3C~273. In our interpretation, a rotating gas disk is affected by the emerging jet due to the associated expanding hot gas cocoon. As the pressurized cocoon is expanding in all directions, it will \textit{predominantly} push the gas down on the approaching jet and up on the receding jet side with respect to our line of sight. This leads to a transverse shock traveling outwards through the disc, which compresses the gas in the pre-shock region where enhanced gas density leads to shorter cooling times and formation of molecular gas. Since the approaching jet is much more prominent it is unclear if a receding jet has a similar power to create a density wave on the opposite side to form molecular gas.}
 \label{fig:3C273_cartoon}
\end{figure*}

The sizes of outflows have been measured by defining a kinematically disturbed region (KDR) either based on a fixed cut in line width, $W_{80}>600\,\mathrm{km\,s}^{-1}$ \citep{Sun:2017}, or in relation to the stellar velocity dispersion, $\sigma_{\mathrm{[OIII]}}/\sigma_*>1$ \citep{Kang:2018}. For 3C~273 we measure $\sigma_*=210\pm10$\,$\mathrm{km\,s}^{-1}$, which is close to the expected value of $\sigma_*\sim$250\,$\mathrm{km\,s}^{-1}$ from the $M_\mathrm{BH}$--$\sigma_*$ relation \citep{Kormendy:2013} adopting $M_\mathrm{BH}\sim5\times10^9\,M_\odot$ \citep{Kaspi:2000}, so that both KDR definitions are comparable. Indeed, previous measurements reveal a tight relation between $L_\mathrm{bol}$ and $R_\mathrm{KDR}$ as shown in Fig.~\ref{fig:ENLR_scaling}. For 3C~273 we measure $R_\mathrm{KDR}=6.1\pm1.5$\,kpc, adopting $W_{80}>600\,\mathrm{km\,s}^{-1}$, significantly smaller than the extrapolated relations proposed by \citet{Sun:2017}, but close to that of \citet{Kang:2018}. Interestingly, the relation by \citet{Sun:2017} for $R_\mathrm{KDR}$ predicts larger sizes than the $R_\mathrm{ENLR}$ at the high-luminosity end. As discussed in \citet{Husemann:2016c} it is possible that beam smearing leads to a systematic over-prediction of KDR sizes. This effect is minimized for 3C~273 due to the high physical resolution and application of the AGN-host deblending.

\subsection{Wind-driven density wave due to expanding jet?}\label{sec:wind_results}
Interactions of jets with the ambient medium have been extensively studied in a wide variety of jet properties from compact to extended system \citep[e.g.][]{Tadhunter:1989,Tadhunter:2000,Odea:2002,Holt:2008,Morganti:2013,Mahony:2016,Villar-Martin:2017,Tremblay:2018,Jarvis:2019,Kolwa:2019}. In the vast majority of these cases, the most extreme  gas motions have been associated with the head of the jets as they push through the ambient gas leaving turbulent gas behind their way. The impact of those outflows have been proposed to either suppress star formation (negative feedback) or even promote star formation (positive feedback). In the first case, the outflow is assumed to disrupt dense cold gas in the host galaxies and drag it outside the host so that less cold gas is able to form stars \citep[e.g.][]{Nesvadba:2006,Schawinski:2009,Fabian:2012}. In the second case, a density wave caused by the expanding shock front may allow to rapidly form molecular gas through enhanced cooling in a compressed dense gas phase. The second process has mainly been theoretically predicted  \citep{Silk:2005, Gaibler:2012, Ishibashi:2012, Zubovas:2013}, but some observational evidence for enhancement in star formation has been observed in the filament of Centaurus A \citep{Crockett:2012}, in companion galaxies impacted by jets \citep[e.g.][]{Croft:2006,Molnar:2017}, and nearby galaxies with strong outflows \citep[e.g.][]{Cresci:2015b,Maiolino:2017}.

3C~273 hosts a powerful jet and a bright optical nucleus, so that enough energy is released to provide strong feedback on the ambient gas. Given that there is no ionized gas detected on large scales associated with the head of the jet suggests that the jet axis is not going directly through the gas disk in the host galaxy. Still, there is a clear alignment of the kinematically disturbed ionized gas on kpc scales.  Here, we refrain from computing gas outflow energetics to distinguish between a mechanically or radiatively-driven outflow, because neither the electron densities nor the actual velocities of individual clouds can be robustly measured with our current data.  However, the well-constrained jet axis about $\sim20^\circ$ away from our line-of-sight \citep[e.g.][]{Stawarz:2004} helps us drawing a potential scenario for the evolution of the outflow. The jet axis provides an independent estimate of the 3D orientation of the primary AGN ionization cone axis which can otherwise be estimated from the ENLR morphology and kinematics \citep[e.g.][]{Fischer:2013}.

Surprisingly, we see preferentially receding gas motion on the approaching jet side and approaching gas motions on the opposite side. This is inconsistent with a purely radiatively-driven outflow which should lead to a radial gas acceleration if the ionization cone is roughly aligned with the jet axis. However, the expanding cocoon around the radio jet will lead to shocks propagating in all directions and create turbulent and bulk motion in the surrounding interstellar medium as it expands with time. Due to the inclination of the  gas disk, the expanding jet-driven hot gas cocoon will accelerate the gas in transverse or even back-flow motion. Indeed, the presence of molecular gas with an arc-like structure just outside the kinematically disturbed region of the outflow may be interpreted as the consequence of a gas density wave generated by the expanding shock perpendicular to the jet. While the molecular gas may also be bound to the remnant of an infalling galaxy, we do not see such a well-defined structure in the ionized gas phase. The jet axis breaks the symmetry and exactly goes through the center of the molecular arc as shown in Fig.~\ref{fig:3C273_overview} panel (e) which is unlikely to be a pure coincidence. The motion of the cold gas is not too much  affected by the outflow yet and predominantly traces the ordered gravitational motion like the kinematically quiescent ionized gas at the same location. Hence, we can estimate the inclination of the undisturbed rotation-dominated gas disk by comparing the dynamical mass with the expected enclosed mass. The stellar mass of 3C~273 is roughly $M_*\sim4\times10^{12}\,M_\odot$ according to the $M_\mathrm{BH}$--$M_{*}$ relation of \citet{Haering:2004}. This leads to an intrinsic rotational velocity of $v_\mathrm{rot}=400\,\mathrm{km\,s}^{-1}$ at 8\,kpc distance whereas the observed maximum radial velocity in the ionized and cold gas phase is about $v_\mathrm{rad}=150\,\mathrm{km\,s}^{-1}$. This implies an inclination of $\sim$20$^\circ$ of the gas disk. 

Our proposed scenario is illustrated as a cartoon in Fig.~\ref{fig:3C273_cartoon}. As the jet has expanded close to the speed of light for $10^5$yr in 3C~273 \citep{Stawarz:2004}, it has already reached beyond the galaxy and is not directly interacting with the gas anymore. However, the continuously expanding cocoon of the jet still affects the gas disk and pushes it in the transverse direction over time.  Those transverse and backflow motions from the expanding cocoon have been observed in hydro-dynamical simulations of expanding jets in a dense gas medium \citep{Mukherjee:2018}. This transverse motion and shock-front generates a symmetric density wave in the pre-shock phase which allows formation of a significant reservoir of cold molecular gas through enhanced cooling. While a counter jet was recently detected in 3C~273 \citep{Punsly:2016} and may also explain the disturbed kinematics on the other side of the AGN, it may be intrinsically less powerful and may fail to produce a shock-front in which molecular gas can form. Since the ionized emission line ratios do not show an enhancement of H$\beta$ with respect to [\ion{O}{3}] at the location of the cold gas suggests that the cold gas is not actively forming stars. However, the $\alpha_{CO}$ factor is unknown and diffuse star formation may not be easily recognized due to the surrounding AGN-ionized region. All those measurements highlight the complex interplay between outflows and the multi-phase medium which depends on a lot of parameters.

While the kinematics may also be interpreted as inflow motion, the velocities of the gas clouds of up to $\pm$800\,$\mathrm{km\,s}^{-1}$ are too high to be driven purely by gravity. Nevertheless, the accelerated gas clouds will contribute to the turbulence in the system and likely enhance the inflow of some material towards the nucleus. Such a chaotic inflow of gas may account for the flicker-noise AGN variability which is predicted by the CCA model \citep{Gaspari:2017} and discovered in 3C~273 over a 80\,yr light-curve monitoring \citep{Press:1978}. A more detailed model of the ionized gas distribution and kinematics to overcome the current limitations of the seeing limited observations will be performed with future high-resolution observations granted with the  new MUSE narrow-field mode.

\subsection{Implications for high-$z$ AGN observations}
It is well known that there is an anti-correlation between the Eddington ratio of an AGN and the strength of the [\ion{O}{3}] line in the NLR \citep[e.g.][]{Boroson:1992,Boroson:2002,Marziani:2003b}. Due to selection effects, high Eddington ratios dominate the luminous AGN population at high redshift and it has been proposed that the NLR is disappearing in these AGN \citep[e.g.][]{Netzer:2004}. Indeed the ENLR on kpc scales remain undetected preferentially around high Eddington ratio AGN in a sample of luminous AGN at $z<0.3$ \citep{Husemann:2008}. 3C~273 belongs to the high Eddington-ratio regime with $L_\mathrm{bol}/L_\mathrm{Edd}\sim0.6$ \citep{Husemann:2013a} and shows a weak [\ion{O}{3}] line in the AGN spectrum. The ENLR was indeed reported to be undetected in initial IFU observations of this source \citep{Husemann:2013a}. We conclude from this work that the high contrast between the luminous AGN and the ENLR has apparently caused previous non-detections. It is simply more difficult to detect the ENLR underneath a beam-smeared point-like AGN signal and requires much higher S/N observations to detect the [\ion{O}{3}] line for a given surface brightness limit. Several AGN surveys at high redshift have attempted to detect the ionized gas around AGN such as KASHz \citep{Harrison:2016b}, WISSH \citep{Bischetti:2017} or SUPER \citep{Circosta:2018}. Our MUSE data provide an ideal data set to empirically constrain the distribution, kinematics and surface brightness of the emission lines around an hyper-luminous quasar. It is therefore instructive to simulate observations of 3C~273 at $z=2.3$ where [\ion{O}{3}] is shifted into the $H$ band.

\begin{figure}
 \includegraphics[width=\hsize]{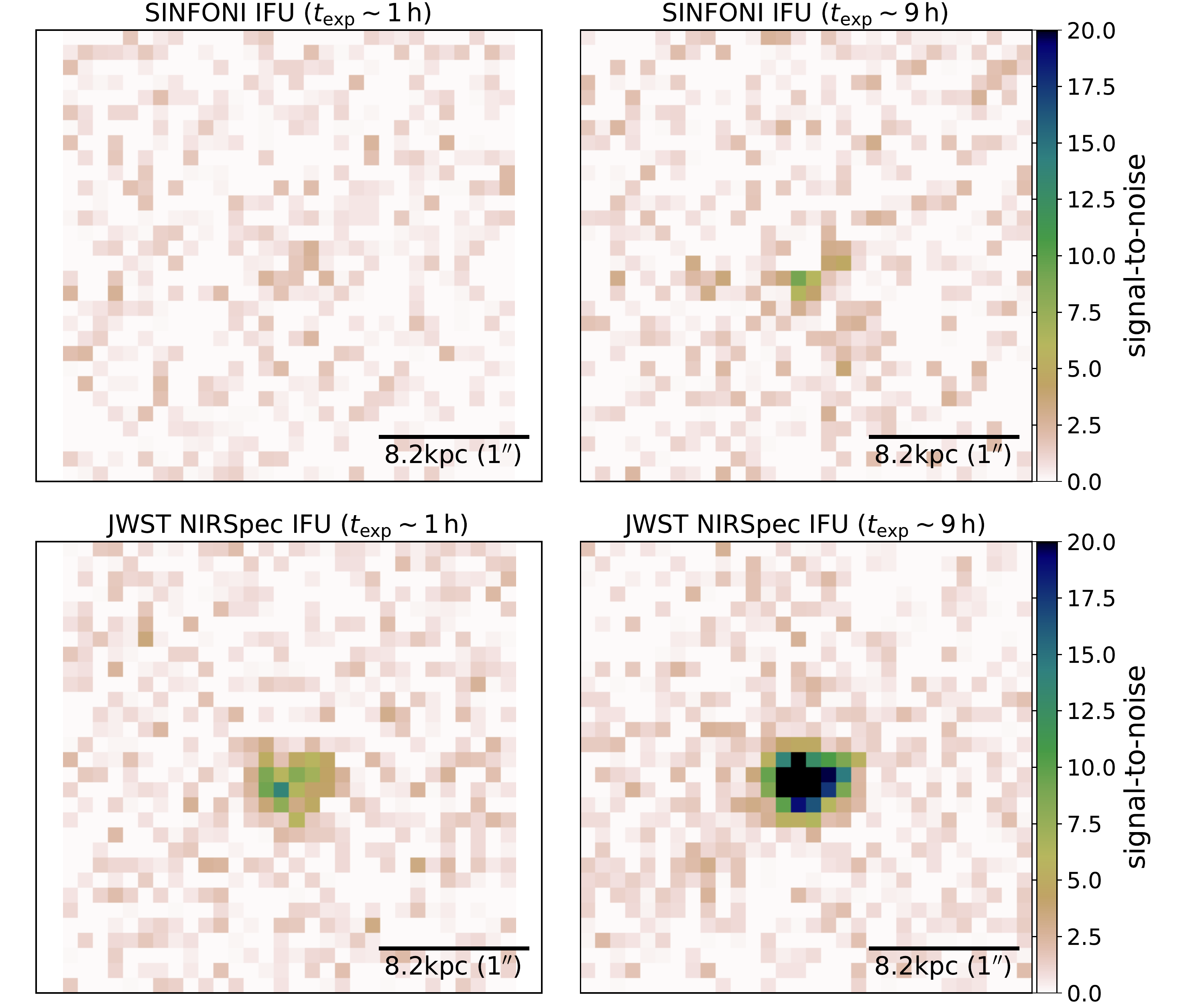}
 \caption{Signal-to-noise maps of reconstructed [\ion{O}{3}] images simulated for SINFONI and JWST observations by redshifting the MUSE observations from $z=0.158$ to $z\sim2.3$ and subtracting the AGN in a consistent way with \QDeb. An empirical noise model (without sky lines) from real observations is used to simulate AO-assisted SINFONI-like observations for 1h and 9h on source integration. The official ETC was used to obtain the noise model for JWST-like observations with the NIRSpec IFU.}
 \label{fig:3C273_comparison_SINFONI}
\end{figure}

The angular scale at $z=2.3$ is 8.2\,kpc/\arcsec\  which is nearly a factor of 3 larger compared to the physical resolution at the redshift of 3C~273. In order to observe such a source at the same spatial resolution as provided by the MUSE data for 3C~273, the angular resolution would need to be 3 times better than our seeing (i.e. $0\farcs23$). This matches the typical angular resolutions  achieved with adaptive optics observations in the $H$ band. Furthermore, the spectral resolution of MUSE is almost the same as the one of the SINFONI $H$-band grating or the high-resolution gratings of NIRspec aboard of JWST. Hence, we have to scale the integrated flux of the MUSE cube after redshifting the spectra by 
\begin{equation}
f = \frac{D_L(z=0.158)^2}{D_L(z=2.3)^2}\frac{(0.158+1)}{(2.3+1)}=5.9\times10^{-4}\label{eq:flux_scale}
\end{equation}
and re-assign the pixel scale from $0.2\arcsec$ to $0.066\arcsec$ or $0.05\arcsec$ for practical reasons to avoid sub-pixel resampling. 

We then simulate realistic SINFONI observations at $z=2.3$ by binning $2\times2$ pixels in the MUSE cube, scaling the fluxes according to Eq.~\ref{eq:flux_scale} and redshifting the spectra. Empirical noise (without sky lines) is added for SINFONI for 1h and 9h of on-source time as directly inferred from real observations with a $0.1\arcsec$ plate scale and a $3''\times3''$ FoV.  We also simulate data for the NIRspec IFU with the same on-source exposure times adopting the noise model of the official ETC \citep{Pontoppidan:2016}. In Fig.~\ref{fig:3C273_comparison_SINFONI} we show the narrow-band images after subtracting the QSO contribution with \QDeb\ in the same way as for the original MUSE data. Ground-based NIR spectrographs such as SINFONI are hardly able to detect the very brightest inner region of the ENLR even after 9h integration time which may explain the low detection rate of kpc-scale [\ion{O}{3}] outflows around optically-bright AGN \citep{Vietri:2018}. NIRSpec will be able to recover the brightest part of the ENLR, but likely miss the full ENLR extent. High-redshift QSO studies therefore need to be aware that a much higher S/N is needed to detect the diffuse extended outflows underneath the highly luminous AGN when designing observations.

\section{Conclusions and outlook}\label{sec:conclusions}
In this article we presented the morpho-kinematic structure of the ionized and molecular gas around the low-redshift hyper-luminous quasar 3C~273 with VLT-MUSE and ALMA in unprecedented detail. We recover a large ENLR with $12.1\pm0.2$\,kpc in size which implies that either the ENLR size-luminosity relation has no break or the break radius is much larger than previously reported. It is possible that CCA plays a relevant role in the formation and properties of 3C~273 ENLR. Within the ENLR, a kinetically disturbed region (KDR) is discovered within the central $\sim$6\,kpc. This KDR is significantly smaller than predicted by literature scaling relations, which are possibly affected by beam smearing. Most strikingly, the KDR is aligned with the radio axis, but preferentially red-shifted on the approaching and blue-shifted on the receding side of the jet contrary to expectations. We interpret this as a signature for an expanding jet-driven hot gas cocoon where the transverse shock is pushing the gas of an inclined gas disk back on the approaching side and up on receding side. This scenario may also explain the presence and morphology of an intriguing molecular gas arc aligned perpendicular to the jet axis.  This feature could be generated by a expanding shock front that leads to a putative density wave where the cooling time is significantly reduced. Whether the molecular gas is forming stars or not is difficult to address with the given MUSE data quality and confusion with the ENLR so that it is unclear whether the molecular gas arc is a long-lived feature or not. Hence, it is unclear if the outflow is leading to positive feedback as predicted for luminous AGN in some theoretical works and simulations. 

We also show that current and upcoming near-IR instruments may be highly limited in recovering the morpho-kinematic structure of the ionized gas on galaxy scales around hyper-luminous quasars at $z>2$ due to the combined effect of surface brightness dimming and the overwhelmingly bright nucleus. The new narrow-field mode of MUSE providing $0\farcs1$ resolution at optical wavelengths will allow us to study the ionized gas around 3C~273 and other luminous AGN in the future, while ALMA can provide details of the molecular gas properties and its excitation conditions. Both facilities together will provide an unprecedented view on the interactions of outflows with the surrounding gas to observationally address the issues raised by positive and negative AGN feedback scenarios. 

\acknowledgements 
\textbf{Acknowledgments}
We thank the referee, Dr. Montserat Villar-Martin, for a careful review and many suggestions that greatly improved the quality of the manuscript. BH acknowledges financial support by the DFG grant GE625/17-1. VNB is grateful to Prof. Dr. Hans-Walter Rix and the Max Planck Institute for Astronomy, Heidelberg, for the hospitality and financial support during her sabbatical stay. MG is supported by the \textit{Lyman Spitzer Jr.}~Fellowship (Princeton University) and by NASA Chandra grants GO7-18121X and GO8-19104X. TAD acknowledges support from a Science and Technology Facilities Council Ernest Rutherford Fellowship.

\noindent This paper makes use of the following ALMA data: ADS/JAO.ALMA\#2015.1.00329.S\\ ADS/JAO.ALMA\#2015.1.00587.S\\ ADS/JAO.ALMA\#2015.1.01012.S\\ ADS/JAO.ALMA\#2016.1.00972.S\\ ADS/JAO.ALMA\#2016.1.01308.S\\ ALMA is a partnership of ESO (representing its member states), NSF (USA) and NINS (Japan), together with NRC (Canada) and NSC and ASIAA (Taiwan) and KASI (Republic of Korea), in cooperation with the Republic of Chile. The Joint ALMA Observatory is operated by ESO, AUI/NRAO and NAOJ.

\end{document}